\documentclass[a4paper]{jpconf}

\usepackage{graphicx}
\usepackage[square,sort&compress,comma,numbers]{natbib}
\usepackage{amsmath}
\usepackage{hyperref}

\begin{document}
\title{Extraction of Polarization Parameters in the $\mathrm{\bar{p} p} \rightarrow \bar{\Omega} \Omega$ Reaction}

\author{E Perotti$^1$}

\address{$^1$ Institutionen f\"or fysik och astronomi, Uppsala Universitet, Box 516, S-75120 Uppsala, SE}

\ead{elisabetta.perotti@physics.uu.se}

\begin{abstract}
A method to extract the polarization of $\Omega$ hyperons produced via the strong interaction is presented. Assuming they are spin 3/2 particles, the corresponding spin density matrix can be written in terms of seven non-zero polarization parameters, all retrievable from the angular distribution of the decay products. Moreover by considering the full decay chain $\Omega\rightarrow \Lambda K\rightarrow p\pi K$ the magnitude of the asymmetry parameters $\beta_\Omega$ and $\gamma_\Omega$ can be obtained. This method, applied here to the specific $\Omega$ case, can be generalized to any weakly decaying hyperon and is perfectly suited for the PANDA experiment where hyperon-antihyperon pairs will be copiously produced in proton-antiproton collisions. The aim is to take a step forward towards the understanding of the mechanism that reigns strangeness production in these processes.
\end{abstract}

\section{Introduction}
The strangeness creation mechanism remains an open question: how are light quarks replaced by heavier ones? Spin observables allow for much deeper insights than just production cross sections. So far several theoretical models have been proposed in order to describe the production of hyperons, often leading to disagreement among each other and with experimental data (see \citep{Ortega:2011zza} and references therein). Colliding protons with antiprotons, one has direct access to hyperon-antihyperon pair production. Hyperons do contain at least one strange quark. In the following we will only focus on the $p\bar{p}\rightarrow\Omega\bar{\Omega}$ case, which involves triple-strangeness baryons and the annihilation of all light quarks in the initial system. It will be studied for the first time at the upcoming PANDA experiment. One of the spin observables that could help us understand how strangeness is created starting from strange-free objects is the polarization of the hyperon. Luckily it can be retrieved experimentally without the need of a polarized beam or target. One exploits the fact that hyperons are weakly decaying -- parity violating! -- particles and therefore their decay products are not isotropically distributed but are emitted along a preferred direction. In brief, from the angular distribution of the decay products one can extract the polarization of the mother hyperon. In weak decays more than one partial wave is populated, i.e.\ both the one violating parity conservation and  the one respecting it. The interference between the two gives rise to the angular dependence in the decay distribution. This is the concrete meaning of the expression (often found in the literature): the weak decays of hyperons are self-analyzing. Exploiting this property we present a method that allows to extract the polarization of the $\Omega$ hyperon.
The Uppsala group is currently performing simulation studies that will complement the material presented here \cite{next}. 
\section{The $\Omega$ spin density matrix}

This section is entirely based on the work done by Doncel \textit{et al.} \citep{Doncel:1973sg, Doncel:1972ez}. The reader is referred to those papers for more details.
The density matrix of a particle with spin $j$ can be written in terms of the polarization parameters $r_{M}^{L}$ as 

\begin{equation}
\rho =  \frac{1}{2j+1}\mathcal{I} + \sum_{L=1}^{2j}\rho^{L}
\label{eq:Q}
\end{equation}

with

\begin{equation}
\rho^{L}=\frac{2j}{2j+1}\sum_{M=-L}^{L}Q_{M}^{L} r_{M}^{L}
\end{equation}
where $\mathcal{I}$ is the identity matrix and $Q_{M}^{L}$ is a set of hermitian matrices. 
The spin density matrix for a particle with $j=3/2$ depends on fifteen polarization parameters $r_{M}^{L}$. However, for a particle produced via the strong interaction, as it is in the $\bar{p} p \rightarrow \bar{\Omega} \Omega$ reaction, parity conservation imposes symmetries on the density matrix \cite{pilkuhn}
so that eight out of the fifteen polarization parameters have to be zero. It follows that the density matrix reduces to

\begin{equation}
\begin{split}
\rho(3/2)= 
\footnotesize{\frac{1}{4}
\left[ \begin{array}{cccc}
1+\sqrt{3}r_{0}^{2} & -i\frac{3}{\sqrt{5}}r_{-1}^{1}+\sqrt{3}r_{1}^{2} & \sqrt{3}r_{2}^{2}-i\sqrt{3}r_{-2}^{3} & -i\sqrt{6}r_{-3}^{3} \\
 &-i\sqrt{\frac{6}{5}}r_{-1}^{3} & & \\ & & &\\
i\frac{3}{\sqrt{5}}r_{-1}^{1}+\sqrt{3}r_{1}^{2} & 1-\sqrt{3}r_{0}^{2} & -i2\sqrt{\frac{3}{5}}r_{-1}^{1}+i3\sqrt{\frac{2}{5}}r_{-1}^{3} & \sqrt{3}r_{2}^{2}+i\sqrt{3}r_{-2}^{3}\\ +i\sqrt{\frac{6}{5}}r_{-1}^{3} & & & \\
 & & & \\
\sqrt{3}r_{2}^{2}+i\sqrt{3}r_{-2}^{3} &i2\sqrt{\frac{3}{5}}r_{-1}^{1}- i3\sqrt{\frac{2}{5}}r_{-1}^{3} &  1-\sqrt{3}r_{0}^{2} & -i\frac{3}{\sqrt{5}}r_{-1}^{1}+\sqrt{3}r_{1}^{2}\\ 
 & & & -i\sqrt{\frac{6}{5}}r_{-1}^{3}\\& & & \\
i\sqrt{6}r_{-3}^{3} & \sqrt{3}r_{2}^{2}-i\sqrt{3}r_{-2}^{3}  & i\frac{3}{\sqrt{5}}r_{-1}^{1}+\sqrt{3}r_{1}^{2} & 1+\sqrt{3}r_{0}^{2}\\& &+i\sqrt{\frac{6}{5}}r_{-1}^{3} & \end{array}  \right]. \mbox{ }\mbox{ }\mbox{ }
}
\end{split}\label{rhopc}
\end{equation}
In other words the $\Omega$ polarization is obtained by measuring the following seven parameters: $r_{-3}^{3}$, $r_{-2}^{3}$, $r_{-1}^{3}$, $r_{2}^{2}$, $r_{1}^{2}$, $r_{0}^{2}$ and $r_{-1}^{1}$. These can be retrieved from the angular distributions of the decay products of the $\Omega$ and its daughter $\Lambda$ hyperon, as will be shown in the following.
We note in passing that for a spin 1/2 hyperon eq.\eqref{rhopc} reduces to a $2\times2$ matrix depending on a single parameter.

\section{Extraction of polarization parameters from decay angular distributions}
\subsection{Single decay: $\Omega\rightarrow \Lambda K$}
The normalized decay $T$ matrix introduced below acts on the density matrix according to $\rho_{\mathrm{fin}} =T\rho_{\mathrm{in}}T^{\dag}$. Conservation of total spin implies that the orbital angular momentum of the final state $\Lambda K$ can be $L=1$ or 2, with parity given by $(-1)^L$. Since this is a parity violating weak decay, both the P- and D-wave are allowed and the components of the $T$ matrix can be written in terms of their amplitudes:
\begin{equation}\label{Tmatrix}
\begin{split}
&\small{T(3/2\rightarrow 1/2,0)=}\footnotesize{\frac{1}{\sqrt{8 \pi}}
\left[ \begin{array}{cccc}
-\sqrt{3}(T_d+T_p) & \frac{1}{2}(T_d+T_p) & -\frac{1}{2}(T_d+T_p)e^{-i\phi} & \sqrt{3}(T_d+T_p)\\ e^{i\phi}\sin{\theta}\cos{\frac{\theta}{2}} & (\cos{\frac{\theta}{2}}+3\cos{\frac{3\theta}{2}})& (\sin{\frac{\theta}{2}}-3\sin{\frac{3\theta}{2}})& e^{-2i\phi}\sin{\theta}\sin{\frac{\theta}{2}}\\ 
 & & & \\
-\sqrt{3}(T_d-T_p) & \frac{1}{2}(-T_d+T_p)e^{i\phi}  & -\frac{1}{2}(T_d-T_p) & \sqrt{3}(-T_d+T_p)\\ e^{2i\phi}\sin{\theta}\sin{\frac{\theta}{2}} & (\sin{\frac{\theta}{2}}-3\sin{\frac{3\theta}{2}})& (\cos{\frac{\theta}{2}}+3\cos{\frac{3\theta}{2}})&e^{-i\phi}\sin{\theta}\cos{\frac{\theta}{2}}\end{array}  \right]. \mbox{ }\mbox{ }\mbox{ }
}
\end{split}
\end{equation}
The decay angles $\theta,\, \phi$ of the $\Lambda$ hyperon are measured in the $\Omega$ rest frame, where the $xz$-plane constitutes the $\Omega$ production plane:
\begin{equation}
\hat{x}=\hat{y}\times\hat{z} \quad\quad \hat{y}=\frac{\vec{p}_\Omega\times\vec{p}_{beam}}{|\vec{p}_\Omega\times\vec{p}_{beam}|} \quad\quad \hat{z}=\hat{p}_\Omega.
\end{equation}
Here we adopt the Jacob-Wick decomposition in partial waves \cite{Jacob:1959at} which refers to helicities. This choice avoids complications when considering also the subsequent decay of the $\Lambda$ hyperon. 
The angular distribution is then obtained by taking the trace of the final state density matrix \cite{koch}:
\begin{equation}
I =\mathrm{Tr} ( \rho_{\mathrm{fin}})=\mathrm{Tr} (T \rho_{\mathrm{in}} T^{\dag})=\mathrm{Tr} (\rho_{\mathrm{in}} T^{\dag}T).
\end{equation}
The normalized P- and D-amplitudes are combined into the following parameters:
\begin{equation}\label{asy}
\begin{split}
&\alpha_\Omega=2 \mathrm{Re}(T_d^{*}T_p) \\
&\beta_\Omega=2 \mathrm{Im}(T_d^{*}T_p) \\
&\gamma_\Omega=|T_d|^{2}-|T_p|^{2} \mbox{ }\mbox{ }\mbox{ }
\end{split}
\end{equation}
which quantify the asymmetries in the decay angular distribution. By construction they fulfill $\alpha^2+\beta^2+\gamma^2=\lvert T_p\rvert^2+\lvert T_d\rvert^2=1$.
The angular distribution then reads:
\begin{equation}\label{310}
\begin{split}
I(\theta,\phi)&=\mathrm{Tr}(\rho(3/2)T^{\dag}T(3/2\rightarrow 1/2,0))\\ &=\frac{1}{4\pi}\bigg[ 1  +\frac{\sqrt{3}}{2}(1-3\cos^{2}{\theta})r_0^2-\frac{3}{2}\sin^{2}{\theta}\cos{2\phi}\,r_2^2 -  \frac{3}{2}\sin{2\theta}\cos{\phi}\,r_1^2 \\ & \quad +\frac{1}{40} \alpha_\Omega \sin
  \theta \Big(8 \sqrt{15} r_{-1}^1 \sin\phi - 
   9 \sqrt{10} r^3_{-1} (3 + 5 \cos2 \theta)\sin\phi\\ &\qquad -
   30 (3 r^3_{-2} \sin2 \phi \sin2 \theta+\sqrt{6} r^3_{-3} \sin3 \phi \sin^2\theta)\Big)\bigg].
\end{split}
\end{equation}
Of the three asymmetry parameters only the $\alpha_\Omega$ parameter shows up in the angular distribution of the $\Omega$ decay, whereas the $\beta_\Omega$ and $\gamma_\Omega$ parameters will appear when looking also into the subsequent decay of the daughter $\Lambda$ hyperon. The asymmetry parameter $\alpha_\Omega$ has been measured experimentally for the $\Omega\rightarrow\Lambda K$  decay ($\alpha_\Omega=0.0180\pm0.0024$  \cite{Olive:2016xmw}), while $\beta_\Omega$ and $\gamma_\Omega$ are unknown. 
Note that all seven non-zero polarization parameters show up in eq.\eqref{310} but only three of them are not multiplied by the parameter $\alpha_\Omega$. For the moment we focus on these ones only. They can be retrieved if the following expectation values are measured:
\begin{equation}\label{3r}
\begin{split}
\langle \sin{\theta_{\Lambda}}\rangle   
&=\int   I(\theta_{\Lambda}, \phi_{\Lambda}) \times
\sin{\theta_{\Lambda}} \,\text{d} \Omega_{\Lambda} 
= \frac{\pi}{32} (8 + \sqrt{3} r_0^2)\\
 \langle \cos{\phi_{\Lambda}}\cos{\theta_{\Lambda}}\rangle   
& =\int   I(\theta_{\Lambda}, \phi_{\Lambda}) \times
\cos{\phi_{\Lambda}}\cos{\theta_{\Lambda}}\,\text{d} \Omega_{\Lambda} 
 =-\frac{3\pi}{32} r_1^2\\
 \langle \sin^2{\phi_{\Lambda}}\rangle   
& =\int   I(\theta_{\Lambda}, \phi_{\Lambda}) \times
\sin^2{\phi_{\Lambda}}\,\text{d} \Omega_{\Lambda}
 =\frac{1}{4}(2+ r_2^2)\\
\end{split}
\end{equation}
where $\text{d} \Omega_{\Lambda}= \sin{\theta_{\Lambda}}\text{d} \theta_{\Lambda} \text{d} \phi_{\Lambda} $ and $\theta_\Lambda\in[0,\pi]$, $ \phi_\Lambda\in[0,2\pi]$. This technique is known as the Method of Moments \cite{frodesen}. We have slightly modified the angle notation ($\theta\rightarrow\theta_\Lambda$,\, $\phi\rightarrow\phi_\Lambda$) to clearly indicate that these angles belong to the first decay, i.e. $\Omega\rightarrow \Lambda K$.
Similar expressions can be obtained also for the remaining four $r$'s but we postpone their derivation to the next section.

\subsection{Full decay chain: $\Omega\rightarrow \Lambda K\rightarrow p\pi K$}
At this point four non-zero polarization parameters remain to be extracted. The reason why we have not done it so far is that we do not want the small parameter $\alpha$ to appear in their expressions. We expect to gain additional information by studying the angular distribution of the subsequent decay of the daughter $\Lambda$ hyperon. This angular distribution can be obtained by letting the $\tilde{T}$ matrix for the $\Lambda \rightarrow p \pi$ decay act on $\rho_{\mathrm{fin}}$, i.e.
\begin{equation}\label{ad}
I  =\mathrm{Tr} (\tilde{T}\rho_{\mathrm{fin}}\tilde{T}^{\dag})=\mathrm{Tr} (\tilde{T} T \rho T^{\dag}\tilde{T}^{\dag})=\mathrm{Tr} (\rho T^{\dag}\tilde{T}^{\dag}\tilde{T}T)
\end{equation}
with $T$ and $\tilde{T}$ referring respectively to the first and second decay. For the $\Lambda \rightarrow p \pi$ weak decay, conservation of total spin implies $L=0$ or 1 in the final state. Since this is a parity violating decay the  $\tilde{T}$ matrix is given in terms of both the S- and P-amplitudes:

\begin{equation}\label{second}
\begin{split}
\small{\tilde{T}(1/2\rightarrow 1/2,0)=}\footnotesize{\frac{1}{\sqrt{4 \pi}}
\left[ \begin{array}{cc}
(\tilde{T}_s+\tilde{T}_p)(\cos{\frac{\theta_\mathrm{p}}{2}}) & (\tilde{T}_s+\tilde{T}_p)e^{-i\phi_\mathrm{p}}(\sin{\frac{\theta_\mathrm{p}}{2}})\\ 
 &  \\
(-\tilde{T}_s+\tilde{T}_p)e^{i\phi_\mathrm{p}}(\sin{\frac{\theta_\mathrm{p}}{2}}) & (\tilde{T}_s-\tilde{T}_p)(\cos{\frac{\theta_\mathrm{p}}{2}})  \end{array}  \right]. \mbox{ }\mbox{ }\mbox{ }
}
\end{split}
\end{equation}
Following our notation, the angles in \eqref{second} carry $p$ as subscript, to indicate that they belong to the second decay, i.e. $\Lambda \rightarrow p\pi$. They are measured in the rest system of the daughter $\Lambda$ hyperon, defined by:
\begin{equation}
\hat{x}'=\hat{y}'\times\hat{z}' \quad\quad \hat{y}'=\frac{\vec{p}_\Lambda\times\hat{y}}{\lvert\vec{p}_\Lambda\times\hat{y}\rvert} \quad\quad \hat{z}'=\hat{p}_\Lambda
\end{equation}
with $\hat{y}$ perpendicular to the $\Omega$ production plane.
The S- and P-amplitudes are combined into the asymmetry parameters according to \eqref{asy}, by simply replacing the D-amplitude with the S-amplitude.
Calculating the trace in \eqref{ad} and integrating over the angles of the first decay one gets the angular distribution of the second decay ($\alpha_\Lambda=0.642\pm0.013$ \cite{Olive:2016xmw}):
\begin{equation}\label{2ad}
I(\theta_\mathrm{p},\phi_\mathrm{p})= \frac{1}{4\pi}\Bigg[1 + \alpha_\Omega\alpha_\Lambda\cos{\theta_\mathrm{p}} -\alpha_\Lambda \left(\sqrt{\frac{3}{5}}r^1_{-1}-\frac{1}{2\sqrt{10}}r^3_{-1}\right)\left(\beta_\Omega\cos{\phi_\mathrm{p}}+ \gamma_\Omega\sin{\phi_\mathrm{p}}\right)\sin{\theta_\mathrm{p}}\Bigg]
\end{equation}
where a linear combination of two polarization parameters shows up. Moreover also $\beta_\Omega$ and $\gamma_\Omega$ appear for the first time.

The full expression for the joint angular distribution $I(\phi_\Lambda,\phi_\Lambda,\theta_\mathrm{p},\phi_\mathrm{p})$ obtained from \eqref{ad} depends on all four angles of the two decays. It contains the asymmetry parameters $\alpha_\Omega,\,\beta_\Omega,\,\gamma_\Omega,\, \alpha_\Lambda$ and all seven polarization parameters. Using the Method of Moments as before one can access the remaining four polarization parameters:
\begin{equation}\label{4r}
\begin{split}
 \langle \sin{\phi_{\Lambda}} \cos{\phi_{\mathrm{p}}} \rangle  
 &=\int    I(\theta_{\Lambda}, \phi_{\Lambda}, \theta_{\mathrm{p}}, \phi_{\mathrm{p}}) \times  \sin{\phi_{\Lambda}} \cos{\phi_{\mathrm{p}}} \,\text{d} \Omega_{\Lambda} \text{d} \Omega_{\mathrm{p}} 
 = -\frac{3 \pi^2 \alpha_{\Lambda}\gamma_{\Omega} r_{-2}^3}{1024} \\
 \langle (3\cos\theta_\Lambda-1) \sin{\phi_{\mathrm{p}}} \rangle  
 &=\int    I(\theta_{\Lambda}, \phi_{\Lambda}, \theta_{\mathrm{p}}, \phi_{\mathrm{p}}) \times  (3\cos\theta_\Lambda-1) \sin{\phi_{\mathrm{p}}}\, \text{d} \Omega_{\Lambda} \text{d} \Omega_{\mathrm{p}} = 
 -\frac{\pi \alpha_{\Lambda}\gamma_{\Omega} r_{-1}^3}{4\sqrt{10}} \\
 \langle \sin{\phi_{\mathrm{p}}} \rangle  
 &=\int    I(\theta_{\Lambda}, \phi_{\Lambda}, \theta_{\mathrm{p}}, \phi_{\mathrm{p}}) \times  \sin{\phi_{\mathrm{p}}}  \,\text{d} \Omega_{\Lambda} \text{d} \Omega_{\mathrm{p}} = 
\frac{\pi \alpha_{\Lambda}\gamma_{\Omega}}{160}\left(-4\sqrt{15}r_{-1}^{1}+\sqrt{10}r_{-1}^{3} \right) \\
 \langle \sin{\phi_{\Lambda}}\cos{\phi_{\Lambda}}  \cos{\phi_{\mathrm{p}}} \rangle   
& =\int  I(\theta_{\Lambda}, \phi_{\Lambda}, \theta_{\mathrm{p}}, \phi_{\mathrm{p}}) \times
 \sin{\phi_{\Lambda}}\cos{\phi_{\Lambda}} \cos{\phi_{\mathrm{p}}}\, \text{d} \Omega_{\Lambda} \text{d} \Omega_{\mathrm{p}}  \\
& = \frac{\pi \alpha_{\Lambda}\gamma_{\Omega}}{640}\left(5\sqrt{6}r_{-3}^{3}+4\sqrt{15}r_{-1}^{1}-\sqrt{10}r_{-1}^{3} \right). 
\end{split}
\end{equation}
Inverting these expressions, the polarization parameters can be extracted.
Moreover the ratio $\beta/\gamma$  can be reconstructed for example from
\begin{equation}
\frac{\beta_\Omega}{\gamma_\Omega}=\frac{ \langle\cos{\phi_\mathrm{p}}\rangle}{\langle\sin{\phi_\mathrm{p}}\rangle}
\end{equation}
or by directly fitting the angular distribution \eqref{2ad} to the function 
\begin{equation}
a+b\sin{\theta_\mathrm{p}}\cos{\phi_\mathrm{p}}+c\sin{\theta_\mathrm{p}}\sin{\phi_\mathrm{p}}
\end{equation}
and then looking at the ratio $b/c=\beta_\Omega/\gamma_\Omega$.
Recalling that the asymmetry parameters satisfy $\alpha^2+\beta^2+\gamma^2=1$ and the value of $\alpha_\Omega$ is known, one can for the first time determine up to a sign $\gamma_\Omega$ and $\beta_\Omega$ which is used for testing CP violation in hyperon weak decays \cite{Donoghue:1986nn,He:1991pf}. It follows that also the last four polarization parameters \eqref{4r} can only be determined up to a sign since $\gamma_\Omega$ appears in their expressions. It is however enough to know their absolute value to get the total polarization of the $\Omega $ hyperon defined by

\begin{equation}\label{tp}
d(\rho)=\sqrt{\sum_{L=1}^{2j}\sum_{M=-L}^{L}(r_{M}^{L})^2} \mbox{ }\mbox{ }\mbox{ }.
\end{equation}

\ack
This work builds heavily on E. Thom\'{e}'s PhD thesis \citep{erik}. I thank K. Sch\"onning and S. Leupold for their guidance through this work. I also acknowledge valuable discussions with A. Pilloni.

\bibliographystyle{iopart-num}

\begin{thebibliography}{10}
\expandafter\ifx\csname url\endcsname\relax
  \def\url#1{{\tt #1}}\fi
\expandafter\ifx\csname urlprefix\endcsname\relax\def\urlprefix{URL }\fi
\providecommand{\eprint}[2][]{\url{#2}}

\bibitem{Ortega:2011zza}
Ortega P~G, Entem D~R and Fernandez F 2011 {\em Phys. Lett.\/} {\bf B696}
  352--358

\bibitem{next}
Ikegami~Andersson W, Perotti E, Leupold S, Sch\"{o}nning K and the
  PANDA~collaboration 2017 {\em In preparation\/}

\bibitem{Doncel:1973sg}
Doncel M~G, Mery P, Michel L, Minnaert P and Wali K~C 1973 {\em Phys. Rev.\/}
  {\bf D7} 815--835

\bibitem{Doncel:1972ez}
Doncel M~G, Michel L and Minnaert P 1972 {\em Nucl. Phys.\/} {\bf B38} 477--528

\bibitem{pilkuhn}
Pilkuhn H 1967 {\em The Interactions of Hadrons\/} (Amsterdam: North-Holland)

\bibitem{Jacob:1959at}
Jacob M and Wick G~C 1959 {\em Annals Phys.\/} {\bf 7} 404--428

\bibitem{koch}
Koch W ed~Nikolic M 1968 {\em Analysis of Scattering and Decay\/} (New
  York-London-Paris: Gordon and Breach)

\bibitem{Olive:2016xmw}
Patrignani C {\em et~al.\/} (Particle Data Group) 2016 {\em Chin. Phys.\/} {\bf
  C40} 100001

\bibitem{frodesen}
Frodesen A~G {\em et~al.\/} 1979 {\em Probability and Statistics in Particle
  Physics\/} (Universitetsforlaget)

\bibitem{Donoghue:1986nn}
Donoghue J~F, Holstein B~R and Valencia G 1986 {\em Phys. Lett.\/} {\bf B178}
  319--323

\bibitem{He:1991pf}
He X~G, Steger H and Valencia G 1991 {\em Phys. Lett.\/} {\bf B272} 411--418

\bibitem{erik}
Thom\'{e} E 2012 {\em Multi-Strange and Charmed Antihyperon-Hyperon Physics for
  PANDA\/} Ph.D. thesis Uppsala University

\end{thebibliography}
\providecommand{\newblock}{}

\end{document}